\documentclass[preprint2]{aastex}

\usepackage{amsmath}

\newcommand{\thetaccd}{\theta_{\rm{ccd}}}
\newcommand{\thetas}{\theta_{\rm{s}}}
\newcommand{\thetaprf}{\theta_{\rm{R}}}
\newcommand{\fsky}{f_{\rm{sky}}}
\newcommand{\ngal}{n_{\rm{g}}}
\newcommand{\sn}{\rm{s/n}}
\newcommand{\sigmacd}{\sigma_{\rm{ch}}}
\newcommand{\drrg}{d_{\rm{rrg}}}
\newcommand{\dcut}{d_{\rm{cut}}}
\newcommand{\dstar}{d_*}
\newcommand{\sigmagamma}{\sigma_{\gamma}}

\newcommand{\Itrue}{I^{\rm{tr}}}

\newcommand{\QE}{\epsilon_{\rm q}}

\newcommand{\D}{~{\rm d}}
\newcommand{\PRF}{{\rm R}}
\newcommand{\PRFCD}{{\rm R}_{\rm ch}}
\newcommand{\PRFSQ}{{\rm R}_{\rm sq}}
\newcommand{\PSF}{{\rm P}}

\newcommand{\vect}[1]{\boldsymbol{#1}}
\newcommand{\meter}{{\rm m}}
\newcommand{\second}{{\rm s}}
\newcommand{\mm}{{\rm mm}}
\newcommand{\nm}{{\rm nm}}

\shorttitle{Pixelation in Weak Lensing}
\shortauthors{High et al.}
\slugcomment{Submitted to PASP}

\begin{document}

\title{Pixelation Effects in Weak Lensing}
\author{F.~William High\altaffilmark{1,2}, Jason
  Rhodes\altaffilmark{3,2}, Richard Massey\altaffilmark{2}, Richard
  Ellis\altaffilmark{2}}
\email{high@physics.harvard.edu }
\altaffiltext{1}{Harvard University Department of Physics, 17 Oxford
  St., Cambridge, MA 02138}
\altaffiltext{2}{California Institute of Technology,  MS 105-24, 1201
  East California Blvd., Pasadena, CA 91125}
\altaffiltext{3}{Jet Propulsion Laboratory, MS 169-506, 4800 Oak Grove
  Drive, Pasadena, CA 91109}


\begin{abstract}

Weak gravitational lensing is a promising probe of dark matter and dark
energy requiring accurate measurement of the shapes of faint, distant
galaxies. Such measures are hindered by the finite resolution and pixel
scale of typical cameras.  On the other hand, as imaging telescopes are
practically limited to a fixed number of pixels and operational
life-span, so the survey area increases with pixel size. We investigate
the optimum choice of pixel scale in this trade-off for a space-based
mission, using the full engineering model and survey strategy of the
proposed SuperNova/Acceleration Probe as an example. Our methodology is
to simulate realistic astronomical images of known shear and to evaluate
the surface density of sources where the shear is accurately recovered
using the Rhodes, Refregier \& Groth algorithm in the context of the
derived dark matter power spectrum. In addition to considering single
exposures, we also examine the benefits of sub-pixel dithering. Although
some of our results depend upon the adopted shape measurement method,
the relative trends, particularly those involving the surface density of
resolved galaxies, are robust. Our approach provides a practical
counterpart to studies which consider the effects of pixelation from
analytic principles, which necessarily assume an idealized shape
measurement method. We find that the statistical error on the mass power
spectrum is minimized with a pixel scale equal to $75$--$80\%$ of the
FWHM of the point-spread function, and that dithering is marginally
beneficial at larger pixel scales.

\end{abstract}



\section{Introduction}
\label{sec:intro}

The gravitational field arising from a massive foreground structure
deflects and distorts the light from distant objects according to the
process of {\it gravitational lensing}
\citep[\textit{cf}][]{bib:blandford, bib:narayan}. Rapid progress has
been made in the accurate measurement of the weakly distorted shapes of
background galaxies in order to determine the distribution of dark
matter in space and time \citep[for reviews
see][]{bib:mellierrev,bib:bartelmannschneider,bib:refregierrev}.  As the
growth of the dark matter power spectrum is also sensitive to both the
gravitational attraction of dark matter and the repulsive effect of dark
energy, weak gravitational lensing is emerging as a very promising
cosmological probe \citep[Dark Energy Task Force,][]{bib:detf}.

Several proposed space-based missions (SNAP,
\citeauthor{bib:snapomnibus}~\citeyear{bib:snapomnibus}; DUNE,
\citeauthor{bib:dune}~\citeyear{bib:dune}) plan to use weak lensing as a
major probe. A key design element is the detector pixel scale,
$\thetaccd$ (arcseconds), which must be optimized in terms of a
trade-off between the fidelity with which the true weak lensing signal
is recovered and the benefits of a large survey area during a fixed
mission lifetime.  In this paper, we investigate the practical
consequences of changing the detector pixel scale, using simulations to
explore empirically the effect on measurements of the cosmological
matter power spectrum.

Extracting the weak lensing signal requires accurate measurements of the
shapes of faint galaxies, which are inevitably degraded by the
convolutions arising from a finite point spread function (PSF) and
detector pixelation.  The PSF of an optical space telescope will be
dominated by diffraction arising from the system pupils, mirror struts,
and other physical structures which perturb incident light. The PSF
blurs the shapes of background galaxies, which need to be deconvolved or
at least corrected. The shape of the PSF, which typically varies as a
function of detector position and time can normally be obtained by
examining point sources or directly by ray-tracing software.

Pixelation similarly degrades faint image shapes.  In the case of a CCD
detector, a photon incident at a photosite, or pixel, photoelectrically
liberates an electron, which is then bound near the CCD surface to be
measured later.  The photosite has finite cross-sectional area and
counts photons that fall anywhere inside its perimeter, averaging over
all sub-pixel scale image features.  Liberated electrons may also
diffuse to the surface of adjacent photosites, an effect termed {\it
charge diffusion}.  Averaging and diffusion will distort the measured
shape of features nearly the size of a pixel or smaller, biasing the
measured weak lensing signal.

While CCD photosites effectively average over small scale image
information, it is nonetheless possible to recover some of the high
frequency information by dithering.  Dithering involves taking multiple
exposures of the same stationary objects so that their sub-pixel
positions in the CCD are different each time. This can be done by
slightly translating the camera between exposures.  Individual dithered
exposures are combined before shape analysis using the \textit{Drizzle}
algorithm \citep{bib:drizzle} or the Fourier techniques of
\citet{bib:lauer1,bib:lauer2}; or during shape analysis by somehow
averaging the measurements.

The \textit{Drizzle} algorithm averages dithers on a pixel grid that can
be finer than $\thetaccd$, resulting in a higher resolution, resampled
image. \citeauthor{bib:lauer1}'s technique accomplishes the same thing
without re-pixelation by combining Fourier transforms from the
individual dithered exposures. We investigate the extent to which
\textit{Drizzling} can recover weak lensing signals from undersampled
data.

Our results are part of a wider investigation of the accuracy of weak
lensing measurements and the effects of various instrumental and
algorithmic parameters.  Our image simulation software is employed by
the Shear TEsting Program\footnote{See
\url{http://www.physics.ubc.ca/$\sim$heymans/step.html}.} (STEP).  The
STEP project comprises independent ``blind'' analyses of simulated data
of various types by international groups with a goal of verifying the
limitations of extant weak lensing measurement algorithms. Simulations
of both ground \citep{bib:step1,bib:step2} and space-based (in prep.)
data have been undertaken, and in particular space STEP will also
explore the effects of pixelation on shear
recovery. \citet{bib:bernstein02} and Bernstein et al.~(in preparation)
perform a similar analysis, but starting from the opposite end of
analytic first principles, studying the irreversible loss of information
during pixelation, even with a perfect shape measurement method. In a
further study, \citet{bib:wlfromspaceII} analyze the effect of varying
the exposure time on weak shear recovery from a SNAP-like
mission. \citet{bib:wlfromspaceIII} study semi-analytically the
trade-off between a wide and deep SNAP survey strategy by looking at the
expected errors on cosmological parameters from weak lensing data.

This paper is organized as follows. In \S \ref{sec:math} we describe a
useful mathematical model of pixelation.  \S \ref{sec:method} introduces
our adopted PSF model (\S \ref{sec:psf}) and simulation software (\S
\ref{sec:simage}) and discusses how we vary the pixel scale (\S
\ref{sec:varying}), our simulated dither and {\it Drizzle} strategy (\S
\ref{sec:dither}), and our shear measurement algorithm (\S
\ref{sec:rrg}).  Results are presented in \S \ref{sec:results}.
Properties of the simulated images are shown in \S \ref{sec:properties},
and then the recovered weak lensing signals are analyzed in terms of the
surface density of useable galaxies (\S \ref{sec:ngal}), the bias
determination (\S \ref{sec:bias}), the standard error of shear
estimation (\S \ref{sec:stderr}), and the predicted error on the matter
power spectrum (\S \ref{sec:deltacl}).  We conclude in \S
\ref{sec:conclusions}.


\section{A Pixelation Model}
\label{sec:math}

Here we introduce the formalism necessary for our analyses.  We define
the true time-independent image $\Itrue(\vect\theta)$ as the
monochromatic photon flux at angular location
$\vect\theta=(\theta_x,\theta_y)$ in the sky, as imaged with a perfect
instrument. A real telescope convolves $\Itrue(\vect\theta)$ with its
``diffraction-pattern'' PSF, $\PSF(\vect\theta)$, which is normalized to
unit integral. In our simulations, we take the PSF to be spatially
independent across the focal plane.

The PSF-convolved image is then pixelated. Integration within a pixel is
equivalent to two distinct operations \citep{bib:lauer1,bib:lauer2}.
First, the PSF-convolved image is again convolved with a {\it pixel
response function (PRF)} $\PRF(\vect\theta)$, which has characteristic
size $\thetaprf$, producing
\begin{equation}
\label{eqn:convolution}
I'(\vect\theta) = \Itrue(\vect\theta) * \PSF(\vect\theta) *
\PRF(\vect\theta),
\end{equation}
where $*$ denotes convolution.  Second, $I'$ is sampled at regular
intervals of size $\thetas$, such that the observed $ij$-th pixel value
is
\begin{equation}
\label{eqn:sampling}
I_{ij} = I'(i\thetas,j\thetas),
\end{equation}
where $i\in\{1,2,\ldots,N_x\}$, $j\in\{1,2,\ldots,N_y\}$, and the
observed image has $N_x\times N_y$ pixels.

It is important to emphasize that pixelation is more complicated than a
re-sampling of the PSF-convolved image.  Photosites integrate photons
incident anywhere within their small yet finite borders, allowing only
for a spatially averaged, or ``binned,'' flux measurement.  The PRF
convolution of Equation [\ref{eqn:convolution}] fully describes this
averaging. CCDs become ideal samplers only in the limit of truly
infinitesimal photosites, in which case the PRF is a $\delta$-function
and the convolution is simply equivalent to the PSF-convolved image.

The ideal CCD with finite photosite size has a single PRF across the
entire array, equal to a unit-normalized square tophat with side
$\thetaprf=\thetaccd$:
\begin{equation}
\label{eqn:sq}
\PRFSQ(\vect\theta) = \begin{cases} \thetaccd^{-2} & \textrm{if
    $|\theta_x|, |\theta_y|<\thetaccd/2$}, \\ 0 &
    \textrm{otherwise}. \end{cases}
\end{equation}
Real PRF's deviate from this idealized case.  For example, imperfect
quantum efficiency, which results in only a fraction of incident photons
being counted, is equivalent to an $\PRF$ normalized to less than unity.
Similarly, variable quantum efficiency across the array is equivalent to
a spatially-varying $\PRF$; and charge diffusion can be regarded as a
PRF with $\thetaprf>\thetaccd$.

We can apply this formalism to consider the performance of single
exposures and dithered images (``dithers'') of various kinds.  A single
exposure is an image $I_{ij}$ with sample spacing $\thetas=\thetaccd$.
A dither refers to the case where the camera is translated by,
e.g.~non-integral pixels $(d_x,d_y)$ delivering an observed image
$I_{i-d_x,j-d_y}$, also with $\thetas=\thetaccd$.  {\it Ideal
interlacing} is a particularly important case consisting of $N \times N$
dithers with $d_x=k/N$ and $d_y=\ell/N$, where $N$ is a positive
integer, and $k,\ell\in\{0,1,\ldots,N-1\}$.  \textit{Drizzle} can
deinterlace such dithers, say from half-$\thetaccd$ shifts ($N=2$), by
applying them to a pixel grid that is twice as fine, yielding an
observed image with $\thetas=\thetaccd/2$.  {\it Drizzling} in general
is an effective ``re-pixelation,'' with its own PRF that is free to be
chosen by the user.  We emphasize that the observed image {\em in all
cases} is convolved with the CCD pixel kernel $\PRF$ with size
$\thetaprf$, and $\thetaprf \neq \thetaccd \neq \thetas$ in general.
That is, even in the limit of infinite, perfectly deinterlaced dithers
with a $\delta$-function re-pixelation kernel, the resultant image is
still convolved with $\PRF$.

We model the PRF as a square top-hat response convolved with an
additional charge diffusion kernel, multiplied by a fraction
$\QE\in(0,1]$ representing quantum efficiency.  The charge diffusion
kernel is taken to be a Gaussian,
\begin{equation}
\label{eqn:chdiff}
\PRFCD(\vect\theta) = \frac{1}{2\pi\sigmacd^2} \exp\left(
\frac{\vect\theta^2}{2\sigmacd^2} \right),
\end{equation}
where $\sigmacd$ is the RMS extent of the diffusion.  Our PRF is then
\begin{equation}
\label{eqn:prf}
\PRF(\vect\theta) = \QE \PRFSQ(\vect\theta) * \PRFCD(\vect\theta).
\end{equation}
The final size of the PRF, $\thetaprf$, is larger than both $\thetaccd$
and $\sigmacd$ due to the convolution.

Following \citet{bib:bernstein02} and assuming $\PSF$ and $\PRF$ are
both space- and time-invariant, we use the associativity of convolutions
to define the {\it effective point-spread function} ePSF, equal to
$\PSF*\PRF$.  Now $\Itrue(\vect\theta)$ is convolved only once, with the
ePSF, and then sampled as in Equation \ref{eqn:sampling}.  Throughout
the paper we distinguish between the PSF, $\PSF$, which only includes
diffraction; the chPSF, equal to $\PSF*\PRFCD$; and the ePSF, equal to
$\PSF*\PRFCD*\PRFSQ$.


\section{Method}
\label{sec:method}

We have developed a pipeline to simulate all relevant steps from the
acquisition of imaging data, through its reduction, to the measurement
of the weak lensing (or ``shear'') signal. We first manufacture
realistic images containing galaxies with a known shear signal, with
various pixel scales, but keeping all other parameters fixed. We then
detect galaxies in the noisy images and measure their shapes. We correct
their shapes for ePSF effects, which are measured from separate,
simulated images of dense stellar fields at each pixel scale. We finally
compare the output, recovered shear measurement to the known input shear
signals. In a parallel pipeline, we also create sets of four shallower
but dithered images at slightly larger pixel scale, which we stack using
{\it Drizzle} to improve the pixel sampling. The interlacing of these
images is optimal, thus providing an optimistic study of how much
resolution could in principle be recovered from a hardware design that
slightly undersamples the PSF.

In the following subsections we describe the key ingredients and
processes in the pipeline.

\subsection{Adopted PSF}
\label{sec:psf}

The assumption of a constant ePSF in space or time isolates the problem
of shear measurement (in which we are primarily interested) without
distractions of PSF interpolation (which is a separable problem that is
being widely discussed elsewhere).

Our ePSF model is based on an early engineering design, called TMA63, of
the SuperNova/Acceleration Probe\footnote{See
\url{http://snap.lbl.gov/}.}  space telescope
\citep[SNAP,][]{bib:snapomnibus}, and accounts for diffraction by the
telescope plus charge diffusion in the CCD. The simulated diffraction
pattern comes from a raytraced model of the SNAP $f/11$ optical system.
It simulates light from an $820\nm$ wavelength point source incident on
a $2\meter$ primary mirror, a secondary supported in front of the
primary by three struts, a folding flat, and a tertiary.  The point
source is taken to be off-axis such that its image appears on the focal
plane $198\mm$ radially away from the optical axis, where the typical
SNAP CCD lies. SNAP plans to employ Lawrence Berkeley National
Laboratory's new high resistivity CCDs, in which charge diffusion
further spreads point source light
\citep{bib:holland,bib:stover,bib:groom}.  We model charge diffusion as
an additional Gaussian convolution with standard deviation
$\sigmacd=4\micron$, the expected level for these CCDs, which have
$10\micron$ wide photosites.  The net chPSF has FWHM of $0.12\arcsec$;
pixelation at the baseline SNAP CCD pixel scale of $0.10\arcsec$ gives a
final ePSF FWHM of $0.14\arcsec$.

\subsection{Simulated Images and Input Shear}
\label{sec:simage}

The image simulation suite of \citet{bib:shapelet_sims},
\textit{Simage}, is our main tool for creating artificial astronomical
images and applying an arbitrary weak shear signal.  \textit{Simage}
uses Shapelets\footnote{See
\url{http://www.astro.caltech.edu/$\sim$rjm/shapelets/} for a Shapelet
analysis package.}, a parametrization of galaxy morphologies as a
weighted sum of a complete, orthonormal set of basis functions
\citep{bib:shapelets1}.

Realistic morphologies are generated by empirically matching the measured 
properties of  actual galaxies in the Hubble Deep Fields
\citep[HDF,][]{bib:hdfn,bib:hdfs}. The HDF galaxy positions, orientations, 
morphologies (plus sizes and magnitudes) are randomized when
generating new images.  In this way realistic magnitude-morphology trends
are produced, though no spatial clustering is imposed and no redshift
information is encoded .  Because the HDF source catalog on which
the simulations are based are already pre-convolved with the HDF PSF, 
they have slightly larger intrinsic sizes than the true galaxy population.
However, this does not affect the process of shear addition and
measurement.

The analytic shapelet models of galaxies are then sheared as
described by \citet{bib:polarshapelets}, including terms up to
fourth order in the shear $\vect\gamma$.  Our input shears $\vect\gamma
= (\gamma_1, \gamma_2)$ range from $-0.05$ to $+0.05$ in steps of
$0.025$ in each component, while the other component is fixed at $0$, viz:

\begin{eqnarray}
\gamma_1 \in \{-0.05, -0.025, 0, 0.025, 0.05 \} \textrm{ and } \gamma_2
= 0;\\ \gamma_1 = 0 \textrm{ and } \gamma_2 \in \{-0.05, -0.025, 0.025,
0.05 \}.
\end{eqnarray}

A total of nine different input shears per pixel scale is thus produced.
We make one image per input shear, uniformly distorting all galaxies in
the field.  The sheared galaxies are convolved with the PSF and transformed 
to real image space, integrating the basis functions analytically within each 
pixel.  This integration is mathematically identical to the convolution with 
the $\PRFSQ$ described in \S \ref{sec:math}.  Finally, photon noise
and a realistic space background signal are added.


\subsection{Pixel Scale}
\label{sec:varying}

The key question motivating this study is what pixel scale most benefits
weak lensing analyses.  To address this we simulate and analyze images
with CCD pixel scales perturbed about a baseline value chosen to be
$\theta_0=0.10\arcsec$.  In practice, pixel scales can be adjusted in
one of two ways and this defines two categories of simulations.

First, we simulate changing the CCD pixel scale $\thetaccd$ by changing
only the physical size of the photosites, labeling this dataset ``Sim
1.''  Second, we simulate changing $\thetaccd$ by adjusting only the
focal length of the telescope, which changes the plate scale, calling
this dataset ``Sim 2.''  While these each linearly perturb $\thetaccd$,
their detailed effects on the final image are distinct and therefore
each merits analysis.

The Sim 1 dataset consists of different simulated images with detector
pixel scales between $0.04\arcsec$ and $0.16\arcsec$ at $0.01\arcsec$
intervals. This ranges from the resolution of the processed, end-product
HDF images, to just larger than the baseline chPSF size. We adopted a
baseline photosite size of $10\micron$, and the different pixel scales
are achieved by varying this between $4\micron$ and $16\micron$ in
$1\micron$ steps. In practice, such changes would need to be made at the
CCD manufacturing stage. Both the diffraction pattern and the baseline
plate scale of $0.01\arcsec/\micron$ are unaffected.  Thus, assuming the
electron diffusion length is constantly $4\micron$ independent of the
photosite size, the different photosite sizes cause no net effect in the
apparent {\it angular size} of the charge diffusion.  The final shape
and angular size of the chPSF, that is before pixelation, are left
unchanged at all $\thetaccd$.  This scheme for pixel scale adjustment
represents a relatively academic exercise into the effect of pixelation
of a fixed PSF.

The Sim 2 dataset emulates the effect of changing the focal length while
keeping the physical size of the CCD photosite fixed, thereby changing
the plate scale and $\thetaccd$ from $0.04\arcsec$, to $0.16\arcsec$.
At longer focal lengths, the Airy disc from diffraction, which is
constant in angular size, grows linearly in microns at the focal plane
while the pixel scale decreases in arcseconds and the charge diffusion
remains at $4\micron$. At these small pixel scales, the diffraction
pattern dominates the chPSF as the first ring is readily apparent.  At
shorter focal lengths, the Airy disc shrinks while the pixel scale
increases.  In this case the charge diffusion dominates, smoothing out
the diffraction features and broadening the chPSF. This is perhaps a
more practical engineering solution to adjusting the pixel scale.

In both cases, the $\thetaccd=0.04\arcsec$ images contain
$4096\times4096$ pixels. These dimensions change linearly with
$\thetaccd$ such that the subtended solid angle per image is constant.
This does not necessarily simulate how an instrument would change its
CCD array size with $\thetaccd$ in practice.  We add about $6000$
galaxies and no stars to each sheared image, extending their
distribution well below the intended detection threshold. Adding noise
to fix the survey depth to $m\simeq27.7$ with $2000\second$ exposures,
we reproduce the number of galaxies useful for weak shear estimation,
$\ngal\approx100$ per arcmin$^2$, found at the baseline pixel scale in
earlier studies \citep{bib:wlfromspaceII}.  For each pixel scale, we
also make one additional image containing only stars, simulating a weak
lensing survey that periodically points at stellar fields to
characterize the ch- and ePSF.

\begin{figure}
\epsscale{1.0}
\plotone{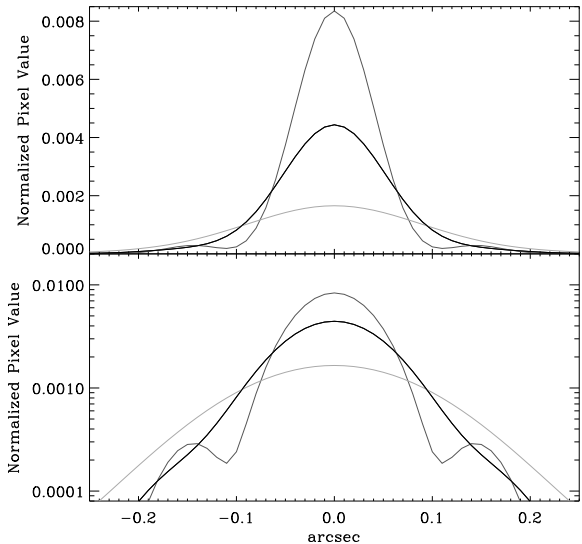}
\caption{ Cross sections through the peak of the simulated point-spread
  function on linear (top panel) and log (bottom panel) scales, with
  components from diffraction and charge diffusion, then finely
  pixelated at $0.01\arcsec$; this is what we term the chPSF.  The
  baseline chPSF (black line) comes from raytracing simulations of the
  current SNAP design, convolved with a 2D Gaussian to approximate
  charge diffusion at the CCD.  In this study we perturb the baseline
  SNAP detector pixel scale of $0.10\arcsec$ in two realistic ways, in
  parallel, which have distinct effects on the chPSF shape.  First, we
  change only the CCD photosite size in microns, which keeps the angular
  size of the charge diffusion kernel constant and therefore leaves the
  pre-pixelated PSF shape unchanged.  Second, we adjust only the focal
  length, which changes the plate scale and thus the angular size of the
  charge diffusion.  In this case, a smaller plate scale reduces the
  effect of charge diffusion over the diffraction pattern in the final
  chPSF (light gray line), while a shorter focal length causes the
  diffraction pattern to dominate (light gray line). }
\label{fig:psfslice}
\end{figure}

\begin{figure}
\epsscale{1.0}
\plotone{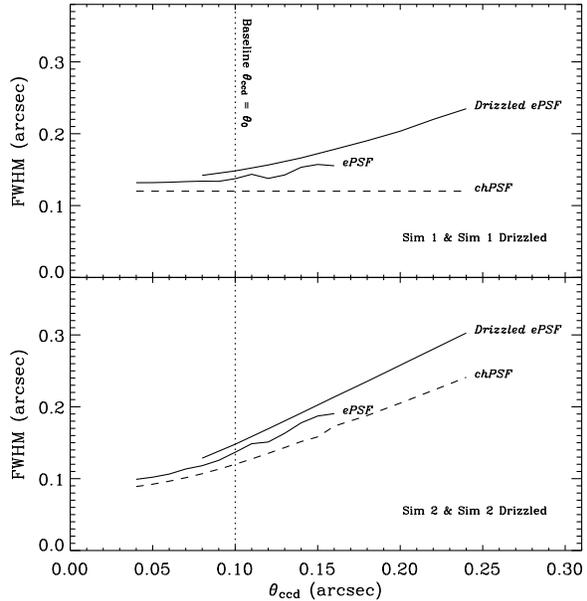}
\caption{ Measured full-width-at-half-maximum values of the ch- and
  ePSFs.  The chPSF in the Sim 1 ({\it Drizzled}) is independent of the
  CCD pixel scale, while the chPSF in the Sim 2 ({\it Drizzled})
  increases in angular size with increased $\thetaccd$.  The plotted
  chPSF FWHM values (dash lines) are measured from highly oversampled
  images using Source Extractor.  Shapelet coefficients are measured
  from the oversampled images, and later re-pixelated in the simulated
  astronomical images at the plotted CCD pixel scales.  The ePSF FWHM's
  shown are averaged from Source Extracted images containing only stars.
  The {\it Drizzled} images are resampled on pixel grids that are twice
  as fine as $\thetaccd$; we measure stars on the {\it Drizzled} images,
  though we plot their CCD pixel scale values instead to allow
  comparison with their true underlying chPSFs.  As this plot shows,
  pixelation increases the apparent size of the point-spread function
  due to the convolution.  {\it Drizzling} introduces an additional
  pixelation, which further increases the ePSF size.  These effects must
  be considered in precise shape analyses such as weak lensing
  requires. }
\label{fig:psfsize}
\end{figure}

Figure \ref{fig:psfslice} illustrates the effect of perturbed CCD pixel
scale on the chPSF profile in the two simulation sets.
The net result is that the underlying chPSF in all Sim 1 images is
constant, while in Sim 2 the chPSF shape and size change with
$\thetaccd$.  Pixelation further distorts the chPSF to produce the ePSF,
which we measure and plot in Figure \ref{fig:psfsize}.
Also plotted in Figure \ref{fig:psfsize} are ePSF sizes after {\it
Drizzling}, which we discuss presently.

\subsection{Simulating and Combining Dithers}
\label{sec:dither}

In addition to the single exposures introduced above, we also create two
additional datasets, labeled ``Sim 1 \textit{Drizzled}'' and ``Sim 2
\textit{Drizzled}'', which simulate dithers that we combine to explore
the extent to which resolution can be recovered, as described in \S
\ref{sec:intro} and \S \ref{sec:math}.  Sim 1 \textit{Drizzled} emulates
Sim 1 in how $\thetaccd$ varies by changing the size of CCD photosites,
and Sim 2 \textit{Drizzled} emulates how Sim 2 changes the pixel scale
by changing the plate scale.

In the \textit{Drizzled} datasets we implement the simplest, ideally
interlaced $2\times2$ dither pattern, where four exposures are taken,
each shifted by exactly half a pixel in orthogonal directions.  These
dithers resample the ePSF-convolved image from Equation
\ref{eqn:convolution} at shifted intevals $(d_x,d_y)$, where
\begin{eqnarray}
  \textrm{Dither 1: }&& (d_x,d_y)=(0,0) \\
  \textrm{Dither 2: }&& (d_x,d_y)=(0.5,0) \\
  \textrm{Dither 3: }&& (d_x,d_y)=(0.5,0.5) \\
  \textrm{Dither 4: }&& (d_x,d_y)=(0,0.5)
\end{eqnarray}
We then deinterlace them using \textit{Drizzle}, placing the four
dithers on a pixel grid that is twice as fine, such that
$\thetas=\thetaccd/2$, and setting \textit{Drizzle} parameter {\sc
pixfrac}=0.5.  We map each dither pixel to one \textit{Drizzled} pixel
with no overlap, such that the noise between adjacent pixels is
uncorrelated. Note that this is an idealized case, impossible to achieve
in practice, and therefore represents an optimistic case study.

All dithers and \textit{Drizzled} images subtend the same solid angle as
in the other two studies.  Each dither is a $500\second$ exposure,
resulting in a $2000\second$ effective exposure after
\textit{Drizzling}.  The CCD pixel scale $\thetaccd$ varies from
$0.08\arcsec$ to $0.24\arcsec$ in steps of $0.02\arcsec$, so the final
\textit{Drizzled} pixel sampling scales, $\thetas$, range from
$0.04\arcsec$ to $0.12\arcsec$.  Table \ref{tab:simsets} summarizes the
four simulation sets .
\begin{deluxetable}{ccccc}
\tablewidth{0pt}
\tablecaption{ Summary of the simulations.
\label{tab:simsets} }
\tablehead{
  \colhead{Sim Set} &
  \colhead{Perturbation\tablenotemark{a}} &
  \colhead{Plate Scale} &
  \colhead{$\sigmacd$\tablenotemark{b}} &
  \colhead{$\thetas$\tablenotemark{c}}
}
\startdata
Sim 1 & Photosite size in $\micron$ & $10\arcsec\mm^{-1}$ & $4\micron =
  0.04\arcsec$ & $\thetaccd$ \\   
Sim 1 {\it Drizzled} &  &  &  & $\thetaccd/2$ \\
Sim 2 & Focal length (plate scale) &
  $10\arcsec\mm^{-1}\times\frac{\thetaccd}{\theta_0}$ & $4\micron =
  0.04\arcsec\times\frac{\thetaccd}{\theta_0}$ & $\thetaccd$ \\  
Sim 2 {\it Drizzled} &  &  &  & $\thetaccd/2$ \\
\enddata
\tablenotetext{a}{What is perturbed in order to change $\thetaccd$.}
\tablenotetext{b}{Standard deviation of the Gaussian charge diffusion
  kernel as a function of perturbed CCD pixel scale.}
\tablenotetext{c}{The sample rate, as a function of perturbed CCD pixel
  scale, of the final images on which the weak lensing analyses are
  applied.} 
\end{deluxetable}

\subsection{Weak Lensing Measurement}
\label{sec:rrg}

We now turn to exploiting the simulated images in order to
recover the weak lensing signal and thereby determine the
surface density of useable galaxies under various assumed
pixel sizes and exposure strategies.

We locate galaxies in the noisy images using Source Extractor
\citep{bib:sextractor} configured to detect all objects as near to the
noise threshold as possible via a Gaussian detection kernel matched to
the known size of the chPSF at each pixel scale.  The size of the
detection kernels are therefore proportional to the chPSF FWHM values
plotted in Figure \ref{fig:psfsize}.  Source Extractor convolves the
images with this detection kernel, blurring features smaller than the
size of the kernel, such as pixel-to-pixel photon counting noise, to
avoid spurious detections.

We adopted the shear measurement method by \citet*[][hereafter RRG]{bib:rrg}
to measure the shapes of detected galaxies. RRG was specifically
developed with space-based weak lensing measurements in mind
\citep{bib:rrg} and has undergone extensive tests on simulations
\citep{bib:leauthaudcosmos} and use on Hubble Space Telescope images
\citep{bib:rrg2,bib:rrg3,bib:rhodesstis,bib:masseycosmos} to constrain
cosmological parameters including $\sigma_8$, the normalization of the
dark matter power spectrum.

RRG is a modification of the KSB+ \citep{bib:ksb,bib:hfks} method which
measures Gaussian-weighted multipole image moments,
\begin{equation}
  J_{ij} = \int \D^2\vect\theta w(\theta) I (\vect\theta)
  \theta_i\theta_j,
\end{equation}
where $w$ is a Gaussian; $i,j\in\{x,y\}$, corresponding to the
orthogonal image coordinates; and $\vect\theta$ is chosen such that the
weighted barycenter is zero.  RRG corrects the galaxy image moments for
ePSF effects using the moments of the measured ePSF. More
advanced shape-measurement algorithms that are being developed,
including the Shapelets-based method of \citet{bib:shapeletswl}, instead
model the chPSF. This requires higher resolution data, but allows
information known {\it a priori} about a regular pixel response function to be
included analytically, whether that be a fixed $\PRFSQ$ or known
pixel-to-pixel variations in the PRF.

RRG forms a measure of ellipticity (in contrast to KSB+), only at the
final stage, after correction of individual shape moments
\begin{equation}
\label{eqn:e}
  \vect e = \frac{(J_{xx}-J_{yy},2J_{xy})}{J_{xx}+J_{yy}}.
\end{equation}
A shear estimator is then formed,
\begin{equation}
  \hat{\vect\gamma} = \frac{\vect e}{G},
\end{equation}
where the shear susceptibility $G$ is a scalar function of higher order
moments of the ensemble of galaxies.

We then remove objects from the shear catalog in a similar manner to
analyses of real data by \citet{bib:masseycosmos,bib:leauthaudcosmos},
and shown by the dashed lines in figure \ref{fig:ngal}.  We first
eliminate the $<1\%$ of galaxies that Source Extractor misclassified as
stars, and also use the Source Extactor $\sn$ outputs \verb|flux_auto|
and \verb|fluxerr_auto| to cut galaxies with $\sn<10$.

\begin{figure}
\epsscale{1.0}
\plotone{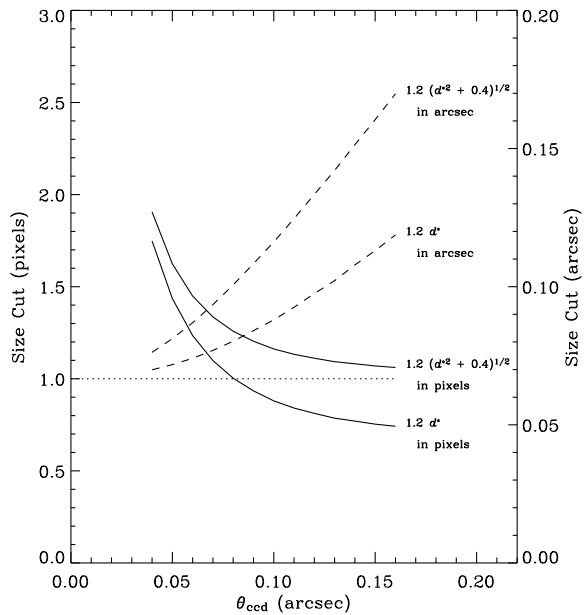}
\caption{ Lower size limit $\dcut$ in pixels (solid, left axis) and
  arsec (dash, right axis) \textit{vs} $\thetaccd$.  (Note the left axis
  doesn't map horizontally to the right axis since $\thetaccd$ is made
  to vary.)  $\dstar$ is the median $\drrg$ measured on high $\sn$ stars
  at each $\thetaccd$.  This study applies cuts with the $0.4$ term,
  which has miminal effect at small $\thetaccd$, but causes the size
  limit to tend toward about one pixel (dotted line) at large pixel
  scales. Plotted here are cuts made in the Sim 2 study, with size cuts
  in other simulation sets behaving similarly.  }
\label{fig:dcut}
\end{figure}

We then calculate the `RRG size', defined using the
uncorrected quadrupole image moments as
\begin{equation}
\label{eqn:drrg}
  \drrg = \sqrt{\frac{J_{xx}+J_{yy}}{2}}.
\end{equation}
We remove large galaxies with $\drrg>2000$, which eliminates fewer than $1$
detected object per arcmin$^2$ in our simulations. And via
\begin{equation}
\label{eqn:dcut}
  \drrg < 1.2 \sqrt{\dstar^2 + 0.4} \equiv \dcut.
\end{equation}
we eliminate galaxies whose size is nearly equal to the
measured ePSF.  The $0.4$ term also eliminates galaxies that are
only a few pixels across, which have $\drrg \approx 1$.  This term is
negligible at small pixel scales, as Figure \ref{fig:dcut} shows,
because the ePSF is large in pixel units.
At large $\thetaccd$, where the ePSF size is nearly one pixel, the $0.4$
term causes $\dcut \to 1$ pixel, also seen in Figure \ref{fig:dcut}.
This cut ensures we measure shear on galaxies that are somewhat larger
than both the PSF and the pixel size, which is the regime shape
measurement is most reliable.  The effect of the lower size cut on the
data is shown with the dashed line below the ellipticity cut in Figure
\ref{fig:ngal}.  We also test applying a lower size limit of $\dcut =
1.2\dstar$ instead of Equation \ref{eqn:dcut}.  This leaves
$\ngal(\thetaccd\lesssim\theta_0)$ unchanged within a few galaxies per
arcmin$^2$ at most, but $\ngal(\thetaccd>\theta_0)$ increases by up to
$\sim10/$arcmin$^2$.  The result that smaller pixel scales always yield
more galaxies, however, is robust.


\section{Results}
\label{sec:results}

After discussing some properties of the simulated images, we explore as
a function of pixel scale: the number of detected and ``useful''
galaxies for weak lensing measurements $\ngal$; the performance of shear
recovery; the sample variance of shear estimators; and the predicted
errors on the dark matter power spectrum.  The number of galaxies for
which shear can be measured is mainly a function of the angular size of
galaxies relative to the size of the PSF.  These results are therefore
likely to be independent of our adopted shear measurement
method. However, the performance of shear recovery and the error on the
power spectrum will depend upon the sophistication of the chosen method,
so those results are likely to improve before the launch of any future
space-based mission.

\subsection{Image Properties}
\label{sec:properties}

\begin{figure}
\epsscale{1.0}
\plotone{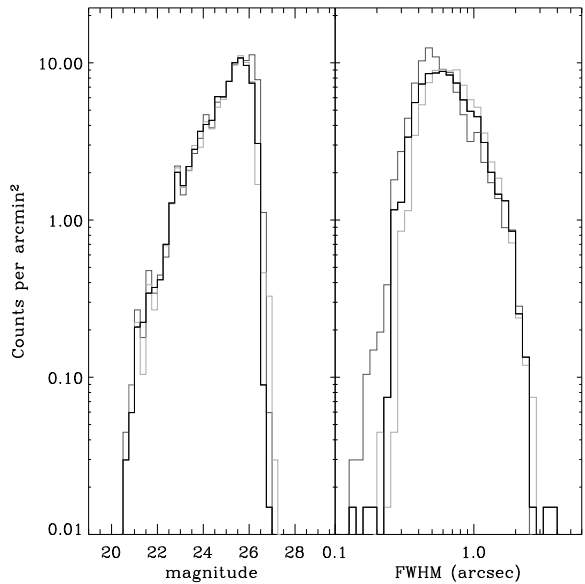}
\caption{ Surface density of galaxies as a function of magnitude (left
  panel) and size (right panel).  These are the galaxies we use to
  measure the shear, which survive various cuts as described in \S
  \ref{sec:ngal}, in Sim 1 images with $\thetaccd=\theta_0$ (black) and
  $\thetaccd= 0.4\theta_0$ (dark gray), and in Sim 1 \textit{Drizzled}
  images with $\thetaccd=2\theta_0$ (light gray). Magnitude bins are
  $0.25$ wide, and size bins are $0.05$ wide in $\log_{10}$ space.
  Magnitude is taken from the Source Extractor
  \texttt{mag\textunderscore auto} output, and FWHM from the
  \texttt{fwhm\textunderscore image} output.}
\label{fig:magsizehist}
\end{figure}

As an illustration, Figure \ref{fig:magsizehist} shows the size and
magnitude distributions of galaxies useful for lensing measurements in
images from Sim~1 at the baseline pixel scale.
The ordinate axis units are counts per arcmin$^2$, and the data are
measured on galaxies after we make cuts, as described in \S
\ref{sec:ngal}.  Our images have a depth of about mag $27.7$, but the
faintest galaxies are not ultimately used.


\subsection{Surface Density of Sheared Galaxies}
\label{sec:ngal}

\begin{figure}
\epsscale{1.0}
\plotone{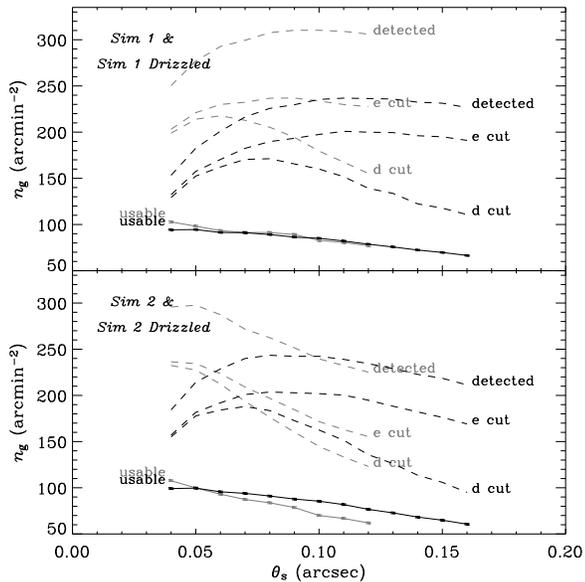}
\caption{ The surface density of galaxies detected and usable for weak
  lensing, $\ngal$, along with counts after different cuts are applied
  to the data in turn, {\it vs} the pixel scale of the reduced images,
  $\thetas$.  The top panel plots data from Sim 1 (black lines) and Sim
  1 {\it Drizzled} (gray lines), and the bottom panel similarly for Sim
  2 and Sim 2 {\it Drizzled}.  Galaxies are detected with Source
  Extractor, and cuts are made based on measured ellipticity (``e
  cut''), size (``d cut,'' see Equations \ref{eqn:drrg} and
  \ref{eqn:dcut}), and $\sn$ (final cut, giving ``usable'' galaxies).
  Error bars are calculated as the sample variance of $\ngal$ between
  the different images at a given pixel scale. }
\label{fig:ngal}
\end{figure}

The surface density of galaxies that survive the various cuts discussed
above is plotted as the solid line in Figure \ref{fig:ngal}.  The uppermost 
dashed lines in each simulation set show the raw galaxy detections using 
Source Extractor\footnote{See
\url{http://terapix.iap.fr/rubrique.php?id\_rubrique=91/}. }, averaged
from all nine sheared images at each pixel scale in each Sim set.  The
surface density of galaxies that survive this cut are plotted in the
next highest dashed line for each Sim set in Figure \ref{fig:ngal}.

As expected, the usable $\ngal$ decreases monotonically in all simulation sets,
indicating that smaller $\thetaccd$ always helps decrease counting noise in
mean shear measurement for a fixed survey area.  Stacking sub-pixel
dithers fully recovers the surface density of galaxies from Sim 1,
despite larger the CCD pixel convolution at pixel scale
$\thetaccd=2\thetas$.  Some galaxies are lost in the Sim 2 {\it
Drizzled} data.  Here the constituent dithers, unlike the Sim 1 sets,
have a larger underlying chPSF than the non-stacked counterparts.  Taken
together, these results indicate that $\ngal$ worsens with increased
charge diffusion, but is independent of the CCD pixel spacing
$\thetaccd$.

We estimate errors on the surface density of usable galaxies, plotted
with error bars in Figure \ref{fig:ngal}, to be the sample variance of
$\ngal$ measurements between each of the nine simulations per pixel
scale.  Errors are of order one galaxy per arcmin$^2$ for all data, which
we consider to be negligible in our analysis.

\subsection{Accuracy of Weak Shear Recovery}
\label{sec:bias}

\begin{figure}
\epsscale{1.0}
\plotone{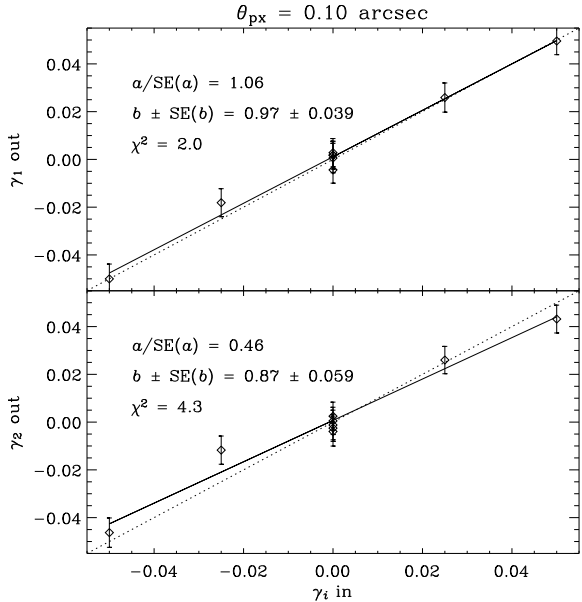}
\caption{Example of known, input shear \textit{vs} measured shear, for
  one set of simulated images.  Points are
  $\langle\hat{\vect\gamma_i}\rangle$, measured from the data.  Error
  bars are the standard error of shear estimators in each image.  The
  dotted lines have slope $1$, representing perfect shear recovery, and
  the solid lines are linear least-squares fits to the data points.
  Also shown are the $y$-intercept $a$ divided by the standard error of
  $a$, ${\rm SE}(a)$, and the slope $b$ plus or minus the standard error
  of $b$, $\rm{SE}(b)$---all outputs from the least-squares
  algorithm. Also shown are the $\chi^2$ values of the fit, which has 7
  degrees of freedom. }
\label{fig:gammaio}
\end{figure}

Figure \ref{fig:gammaio} illustrates the accuracy with which we recover
shear measurement with RRG for one typical set of simulated images.

We define `accuracy' to be the closeness of the measured value to that
originally input. In all cases, the shear recovery is well-fit by a
linear model thereby justifying the catalog cuts described in
\S\ref{sec:rrg}. Relaxing the cuts and using smaller or fainter objects
introduces systematic effects that cannot be corrected with RRG. Again
in all cases, the $y$-intercepts (``additive shear residual'') of the
shear recovery are consistent with zero and accordingly set explicitly
so in subsequent analyses.

\begin{figure}
\epsscale{1.0}
\plotone{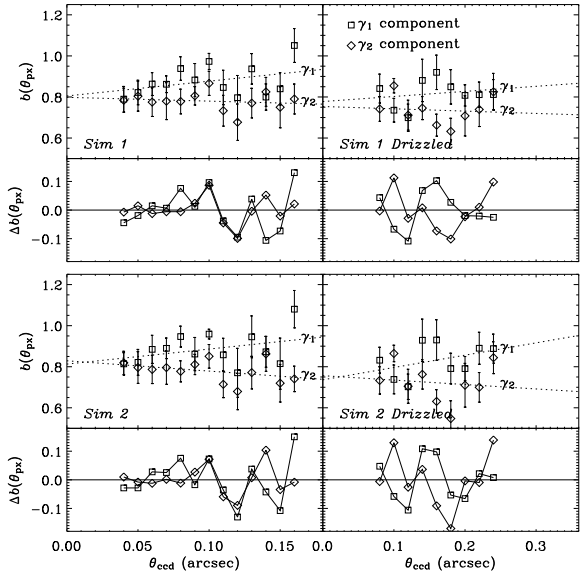}
\caption{ The RRG shear estimator bias $\vect b$, and deviation from the
  least-squares line fitted to each component $\Delta b$, as a function
  of pixel scale $\thetaccd$.  Squars indicate the $\gamma_1$ component
  of shear, relative to the pixel grid, and diamonds the $\gamma_2$
  component.  Error bars show the standard error on slope measurement
  obtained from the least-squares fits to data like those show in
  figure~\ref{fig:gammaio}.  The bias in each component of shear is
  different, and varies with pixel scales.  Extrapolating the pixel
  scale to zero appears to give a bias of about $0.8$, so pixelation is
  likely not the cause of the bias.  {\it Drizzling} to smaller pixel
  scales does not remove the bias or the trend toward $0.8$.  }
\label{fig:cf}
\end{figure}

On the other hand, the best-fit slopes (``multiplicative shear bias'') 
are systematically smaller than unity by roughly $20\%$. This effect 
has been known for some time as a limitation of KSB+ methods 
\citep[\textit{eg}][]{bib:bacon01,bib:step1,bib:step2},
and it has been speculated to arise as a result of pixelation. The
population bias has appeared to be robust to effects like galaxy
morphological type, and we assume in all subsequent sections that it
could, in practice, be determined to arbitrary accuracy using
simulations. We therefore take
\begin{equation}
  \vect b \equiv \frac{\rm{bias}}{\vect\gamma} + 1,
\end{equation}
where, conventionally,
\begin{equation}
  \rm{bias} \equiv \langle \hat{\vect\gamma} \rangle - \vect\gamma.
\end{equation}
to be the true value with negligible error, and correct for the bias
using values from the line fitted as a function of pixel scale in
figure~\ref{fig:cf}.  We call $\vect b$ the bias for brevity.

\subsection{Precision of Weak Shear Recovery}
\label{sec:stderr}

\begin{figure}
\epsscale{1.0}
\plotone{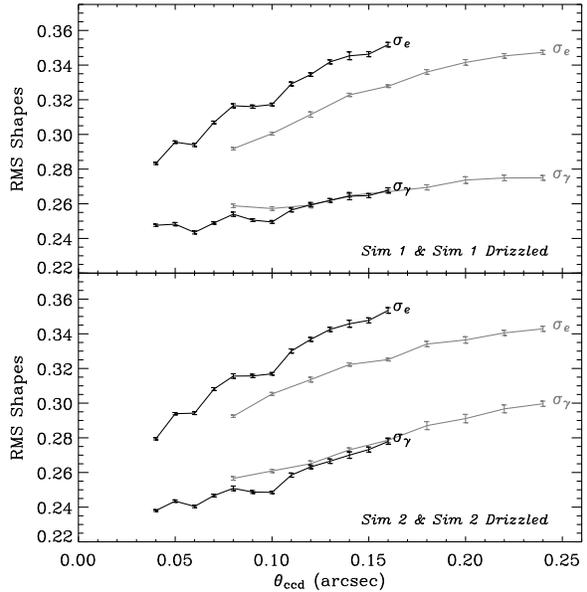}
\caption{ RMS galaxy shapes in single exposure (black lines) and {\it
  Drizzled} (gray lines) simulations.  The RMS of the shear is
  calculated as $\sigmagamma = \sqrt{ \sigma_{\gamma_1}^2 +
  \sigma_{\gamma_2}^2 }$, where $\sigma_{\gamma i}$ is the RMS of the
  bias-corrected shear estimators.  Errors are calculated as the sample
  variance of the RMS shapes between different images at a given pixel
  scale.  The RMS of the galaxy ellipticities are calculated similarly
  from uncorrected ellipticities.  }
\label{fig:sigmagamma}
\end{figure}

Weak lensing induces changes of only a few percent in galaxy
ellipticities, but the RMS of the ellipticity distribution of faint galaxy populations are
about $30\%$, \citep[e.g.][]{bib:leauthaudcosmos}.  Figure
\ref{fig:sigmagamma} shows the standard deviation of ellipticities
$\sigma_e$ and shear estimators $\sigmagamma$, where $\sigma_e$ is
defined analogously to
\begin{equation}
\label{eqn:sigmagamma}
\sigmagamma^2 = \sigma_{\gamma_1}^2 + \sigma_{\gamma_2}^2,
\end{equation}
where
\begin{equation}
\sigma_{\gamma_i}^2 = \langle (\hat\gamma_i - \langle \hat\gamma_i
\rangle)^2 \rangle.
\end{equation}

The measured RMS ellipticity increases monotonically with
$\thetaccd$, meaning larger pixels always make shape measurements
noisier, an undesirable effect for weak lensing analysis.  The ePSF
``dilutes'' the shear signal in galaxies, and RRG is shown here to
reverse this dilution at all pixel scales by decreasing the RMS of
shapes, as expected.  The near coincidence of the non-{\it Drizzled} and
{\it Drizzled} $\sigmagamma$ lines suggests that the RMS shear depends
most on the CCD pixel spacing $\thetaccd$, and whether or not ideal
deinterlacing is performed.  This is, in fact, the ideal situation: the
best image resolution we can have is fundamentally limited by (in
addition to the diffraction) the CCD pixel response---that is, the ePSF.
Ideal deinterlacing increases the sample rate of a given
object, which is why $\sigma_e$ decreases when {\it Drizzling}.  This
plot shows that RRG recovers the underlying, chPSF convolved shape
information {\it from the ensemble} even without {\it Drizzling}.

\begin{figure}
\epsscale{1.0}
\plotone{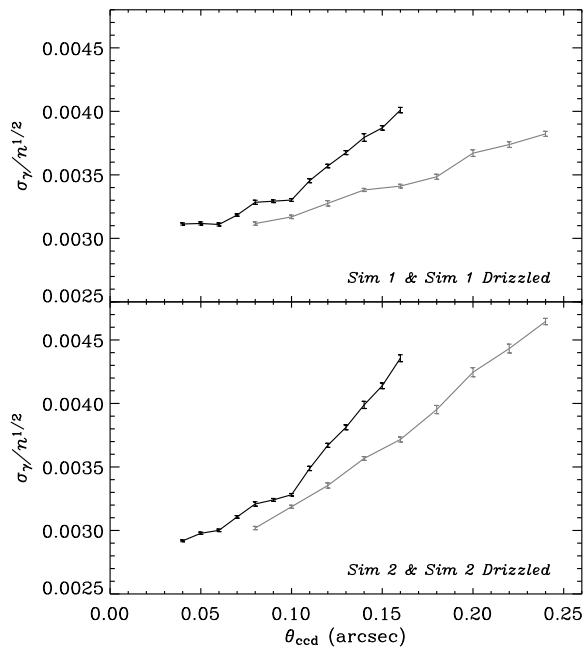}
\caption{ The standard error of the shear, $\sigmagamma/\sqrt{n}$ in the
  single exposure (black lines) and {\it Drizzled} (gray lines)
  simulations.  Error bars are calculated as the sample variance of the
  standard error between the different images at a given pixel scale.
  This quantity combines the method-independent $\ngal$ and the more
  method-dependent $\sigmagamma$ into an estimate of the precision of
  our shear estimation.  This is the figure of merit showing how
  precisely we can measure shear as a function of $\thetaccd$, if the
  survey area is fixed.  }
\label{fig:stderr}
\end{figure}

By assuming uncorrelated shapes in galaxy populations \citep[for a
discussion, see][]{bib:hirataseljak}, this noise can be reduced by
measuring many galaxies and applying Poisson statistics.  An ability to
use more galaxies per arcmin$^2$, $\ngal$, improves the precision of
shear recovery, which is best quantified by the sample variance of shear
estimators, or its square root $\sigmagamma/\sqrt{n}$. This is shown as
a function of pixel scale in Figure \ref{fig:stderr}. This reiterates
our result that smaller pixel scales always improve the precision of
weak shear estimation. However, smaller pixel scale face diminishing
returns: the loss of precision when increasing the pixel scale by some
amount is greater in magnitude than the gain in precision of decreasing
the pixel scale by the same amount.  {\it Drizzling} decreases the error
somewhat, thanks entirely to the behavior of $\ngal$ with $\thetaccd$.
The shear errors from the Sim 2 sets degrade faster with increased
$\thetaccd$. This indicates that charge diffusion, which dominates
larger $\thetaccd$ in the Sim 2 sets, should be minimized in addition to
$\thetaccd$.

\begin{figure}
\epsscale{1.0}
\plotone{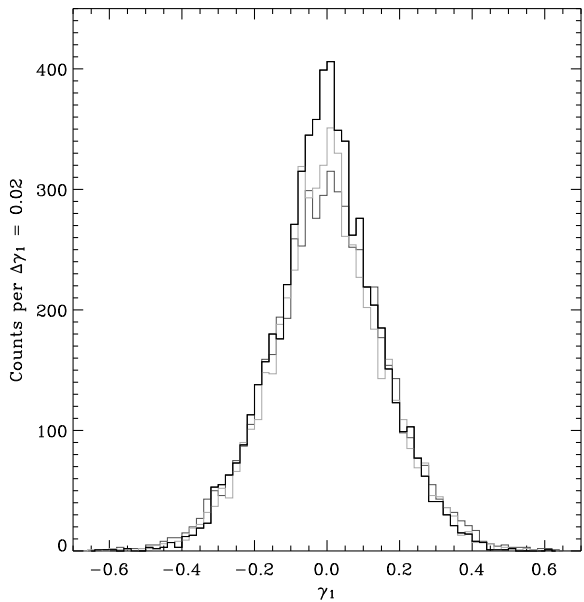}
\caption{ The distribution of shear estimators $\hat\gamma_1$ obtained
  from Sim 1 images with $\thetaccd=0.10\arcsec$ (black) and
  $\thetaccd=0.04\arcsec$ (dark gray), and Sim 1 \textit{Drizzled}
  images with $\thetaccd=0.20\arcsec$ (light gray) pixel scales.  Each
  sample comes from all images at the quoted pixel scales, which have
  different input shears applied.  The mean of the input shears at each
  pixel scale, however, is zero, so these distributions are themselves
  centered around zero. }
\label{fig:gammahist}
\end{figure}

\begin{figure}
\epsscale{1.0}
\plotone{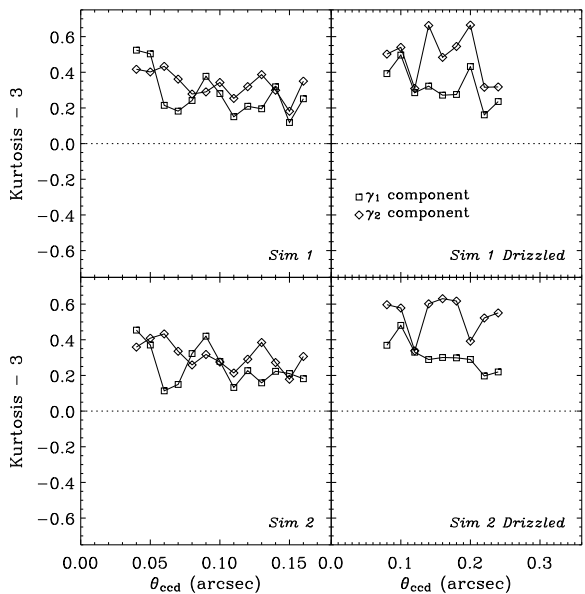}
\caption{ Kurtosis of shear estimators minus $3$ as a function of pixel
  scale.  The kurtosis is calculated as $\langle(\hat\gamma_i -
  \langle\hat\gamma_i\rangle)^4\rangle/\sigma_{\gamma_i}^2$, where
  $\sigma_{\gamma_i}$ is the variance of $\hat\gamma_i$.  The
  $\hat\gamma_i$ are measured from all $9$ images at each $\thetaccd$.
  $3$ is the kurtosis of a Gaussian distribution.  All data here are
  greater than $0$, showing that the distribution of measured shear
  estimators has a sharper peak and more extended wings than a Gaussian.
  }
\label{fig:kurt}
\end{figure}

One potential concern is the observation by Kaiser (2000) that the
distribution of shear estimators in practice is not Gaussian, and that
its extended wings may even make its second moment infinite. This is not
surprising because the shear estimator involves a ratio of two noisy
quantities. Figure \ref{fig:gammahist} shows the distribution of RRG
shear estimators measured in all $9$ images at different pixel scales,
and Figure \ref{fig:kurt} shows the kurtosis of those distributions. The
distributions are indeed non-Gaussian. However, there does not appear to
be any significant degradation in this sense at larger pixel scales. The
skewness of the distributions is also consistent with zero at all pixel
scales.

\subsection{Cosmological Implications}
\label{sec:deltacl}

As we have already seen, the {\it quality} of shear measurements is always
improved with small pixel scales. However, for a mission with a fixed
lifetime, larger pixels would allow a linear increase in the total
survey volume, and a corresponding decrease in sample (or ``cosmic'')
variance errors. These two effects combine in a measurement of the dark
matter power spectrum from cosmic shear, which would have a total
statistical error \citep[\textit{cf}][]{bib:wlfromspaceIII}
\begin{equation}
\label{eqn:deltacl}
\Delta C_\ell=\sqrt{\frac{2}{(2\ell+1)\fsky}} \left( C_\ell +
\frac{\sigmagamma^2}{2\ngal} \right),
\end{equation}
where $C_\ell$ is the power, and $\ell$ is a multipole.  The fraction of
sky surveyed is
\begin{equation}
  \label{eqn:fsky}
  \fsky(\thetaccd)=f_0\frac{\thetaccd^2}{\theta_0^2},
\end{equation}
where $f_0$ is the baseline fraction, $\theta_0$ the baseline pixel
scale, and $\fsky$ at the largest simulated pixel scale is assumed to be
still smaller than the observable sky.  This introduces a tension
between weak lensing precision and cosmic variance. We use the overall
error $\Delta C_\ell$ as our final figure of merit.

\begin{figure}
\epsscale{1.0}
\plotone{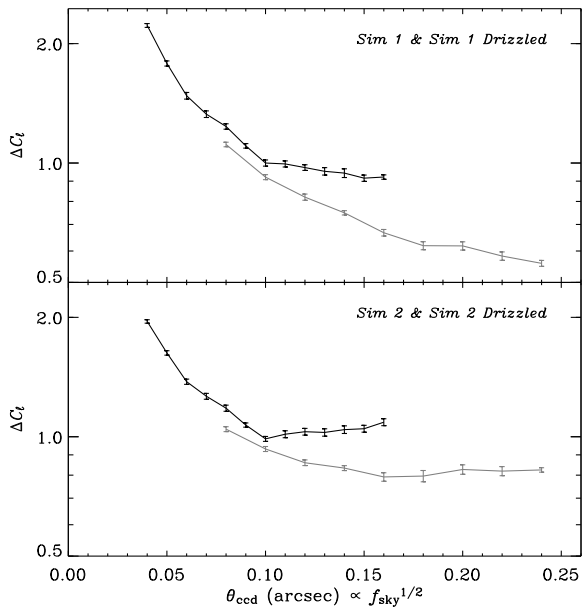}
\caption{ The contribution of shear sample variance to the predicted
  error on the matter power spectrum in single exposure (black lines)
  and {\it Drizzled} (gray lines) simulations.  This assumes the number
  of pixels in the focal plane is fixed when perturbing $\thetaccd$,
  such that the survey area scales as in Equation \eqref{eqn:fsky}.  The
  ordinate axis scale is $\log_{10}$, normalized to the Sim 1 error
  value at the baseline pixel scale, $\thetaccd=0.10\arcsec$.  }
\label{fig:fom}
\end{figure}

The measured values of the second term ({\it i.e.} in the limit of small
$C_\ell$) are shown in Figure \ref{fig:fom}. The ordinate axis uses a
logarithmic scale, in units of the baseline survey because the absolute
values are heavily dependent on the the survey area, multipole $\ell$,
number of galaxies, \textit{etc}.  Cosmic variance dominates errors at
small $\thetaccd$, and increased sample variance in shear estimators
takes effect at large $\thetaccd$.  The error flattens out at larger
pixel scales, affording some freedom in choose the pixel scale, in which
case the smaller pixel scales are clearly preferred for reasons of
caution.

Our idealized \textit{Drizzling} decreases the error, thanks entirely to
the behavior of $\ngal$ with $\thetaccd$, indicating larger CCD pixel
scales are acceptable if perfect deinterlacing is possible and charge
diffusion remains fixed.  It is important to note that larger
$\thetaccd$ degrades $\ngal$, RMS shapes, and ultimately shear errors.
As expected, charge diffusion dominates large at $\thetaccd$ in Sim 2
sets.

Increasing the survey lifetime while fixing the depth would increase the
nominal survey area $f_0$, and thus shifts all curves down uniformly on
a $\log_{10}$ scale; however, this may increase the range of angular
scales probed ($\ell$) and thus change the overall behavior of Equation
\eqref{eqn:deltacl} with $\thetaccd$.


\section{Discussion}
\label{sec:conclusions}

To explore the effects of pixelation on galaxy shape measurement, we
have realistically simulated weakly sheared galaxies at a range of pixel
scales.  We have examined two different ways in which a future
weak lensing space mission could alter the pixel scale, and both non-dithered
and dithered exposure strategies, for a total of four different
simulation sets. We have then emulated the process of shear measurement
that is applied to real data. We have finally combined the observed
surface density of galaxies, statistics of the shear estimators, and our
privileged knowledge of the true input signal, to arrive at figures of
merit for the pixel scale.

We find that, as expected, smaller pixel scales consistently give improve the {\it
quality} of shear recovery and tighten constraints on the dark matter
power spectrum for a hypothetical survey of fixed area.  Ideal
deinterlacing gives a further improvement.  On the other hand, when
the survey area is permitted to change with the pixel scale according to larger
or smaller detectors, we find that larger pixel scales minimize
statistical errors on the measured matter power spectrum because the
survey area increases as $\thetaccd^2$.  These errors flatten out
somewhat above $\thetaccd=0.09\arcsec$.  Considering both
situations of fixed and variable survey area, the best $\thetaccd$ would
be the smallest allowable by the projected dark matter constraints,
which we find to be about $0.75$--$0.80$ the FWHM of the charge
diffusion convolved diffraction pattern, chPSF.

We used a current-generation shear measurement method as the basis of our study---a
snapshot of available technology. Better methods are certainly needed to
fully exploit the ambitious future surveys now being planned. We can
speculate that these may either be better able to cope with poor
resolution, or (more likely, since information is irrevocably lost
during pixelation) require smaller pixels to overcome systematic floors
revealed by the lowering of statistical errors. This is being suggested
by the variable bias in current results, which is now well documented
but poorly understood. It even has to be argued whether the best use of
an expensive space mission would be to minimize statistical errors on an
isolated measurement of the matter power spectrum. By imaging a smaller
region, but with a higher density of useable galaxies, and smaller
errors, a mission could alternatively be used to obtain the phase
information needed for maps or for higher order correlation functions;
or even to calibrate the shape measurement of larger, ground-based
surveys. As shown by our results, this approach would prefer smaller
pixel scales.

Two important simplifications were imposed on our pipeline. Firstly, we
allowed no temporal or spatial variation in the PSF, and a created a
comfortably large number of fake stars to characterize the PSF shape. As
demonstrated in \citet{bib:acspsf}, pixelation especially adds noise to
peaky objects like a diffraction-limited PSF. A typical survey region is
likely to lie at high galactic latitude. If sufficiently bright stars
cannot be imaged within the time taken for the PSF to vary, noise in the
measurement of PSF shapes (which we have not considered) could
potentially dominate that in galaxy shapes. Secondly, real dither
strategies never provide perfect interlacing.  For example, optical
distortions differentially alter the spacing of the pixel grid in
different places, effectively causing $\thetaccd$ to be a function of
position in the focal plane. Consequently, {\it Drizzling} real images
correlates noise between adjacent pixels because it must average nearby
pixel values. Correlated noise hinders both object detection and shape
measurement, so {\it Drizzling} is detrimental to weak lensing. The
dithering implemented here is therefore idealized, and provides a
best-case scenario.

With these caveats in mind, our approach has provided a practical
analysis that is achievable with existing methodology. It is
complementary to studies starting from analytic first principles and
assuming the existence of a perfect shape measurement method. In
practice, our result on the optimum pixel scale should sensibly be
considered as an upper limit, pending future developments in shape
measurement methodology.

\acknowledgments

We are grateful to Steve Kent for providing raytracing software
configured for the SNAP design.  We also thank Alex Refregier, Dave
Johnston, Matt Ferry, Gary Bernstein, Mike Jarvis, Molly Peeples, Chris
Stubbs and Adam Amara for useful discussions.  The Parallel Distributed
Systems Facility\footnote{\url{
http://www.nersc.gov/nusers/resources/PDSF/}}, a Linux cluster run by
the Department of Energy's National Energy Research Scientific Computing
Center, made our large scale simulations and analysis possible: we
particularly thank Iwona Sakrejda for consultation.  This research was
supported in part by a 2004 Caltech Summer Undergraduate Research
Fellowship and Department of Energy grant DE-FG02-04ER41316.



\end{document}